# Isolation solution for extreme environmental vibrations for quantum-enabling cryogenic setups installed on raised frames

Jonah Cerbin and Ilya Sochnikov[1]

*Physics Department, University of Connecticut, Storrs, Connecticut 06269, USA*

## Abstract

Cryogenic quantum sensing techniques are developing alongside the ever-increasing requirements for noiseless experimental environments. For instance, several groups have isolated internal system vibrations from cold heads in closed-cycle dilution refrigerators. However, these solutions often do not account for external vibrations, necessitating novel strategies to isolate the entire cryogenic systems from their environments in a particular set of raised cryostats. Here, we introduce a dual-stage external active vibration-isolation solution in conjunction with a closed-cycle dilution refrigerator that isolates it from the environment. This dual stage includes two sets of active attenuators and a customized steel tower for supporting experimental probes at heights of 3 m from the floor. Both stages achieve 20-40 dB of attenuation with the active systems engaged, corresponding to levels of vibration in the VC-G range (a standardized Vibration Criterion appropriate for extremely quiet research spaces) on the cryostat's room temperature baseplate and the steel tower. Our unique vibration isolation solution therefore expands the applications of modern cryogenic equipment beyond exclusively quiet specialty buildings, rendering such equipment suitable for interdisciplinary, open-floor research centers.

## Introduction

As low-temperature measurement devices advance [1] and become more sensitive [2–9], noise due to vibrations has become of increasing concern [10,11]. Closed-cycle realization of SQUID microscopy [12,13], scanning tunneling microscopy [14,15], and other scanning probe techniques [2,16] demand a high level of precision, but internal and external sources of vibration remain a challenge [17]. Overall, more advanced cooling systems that are replacing conventional liquid helium technologies (due to helium unavailability and high pricing), such as those with a pulse tube, present an added element of internal vibrations. While dry systems such as dilution refrigerators are becoming more sophisticated and popular, a typical installation may require substantial building modifications, e.g. constructing a floor pit; external vibrations may not be handled well, especially as various peripheral probes and devices, such as optical setups, are added. Installing a system without building modifications, e.g. on a tall frame, may further amplify external vibrations.

Many efforts for vibration isolation in closed-cycle cryostats focus on reducing vibrations internally, on the cold plates, with methods that use a tactic based on a cryogenic mechanical low-pass filter with a very low cutoff frequency close to the operating frequency of a cryogenic cold head, which is typically 1.3-1.4 Hz, complicating the design and construction of these low-pass filters [2,18–21]. Other tactics are based on designing the relevant parts to have resonances at higher frequencies; with stiff elements [22], the frequency of the low-pass filter stage can be increased [14,15], making design less challenging.

External vibrations could be an even larger concern. Although internal isolation methods are generally effective, they may become compromised if the externally induced vibrations of the cryostat are very

---

[1] Correspondence to ilya.sochnikov@uconn.edu.

strong—a result of parasitic resonances in supporting frames or environmental sources such as busy roads or building resonances. External vibrations, if large, may even disrupt circulation of the $^3$He-$^4$He mixture, resulting in unstable mK-range temperatures.

One particular difficulty is the required height clearance of such cryostats, which can be as long as 3-4 m when top- or bottom-loading mechanisms are involved. When there is a noisy basement or when the cryostat cannot be lowered below floor level, raised frame structures must be used. Off-the-shelf commercial products do not provide an adequate solution for isolation when they are installed on simple tall pedestals or columns, mainly because simple tall vertical supports are not rigid enough—particularly in the horizontal plane. The system therefore retains very low resonance frequencies, following the behavior of an inverted pendulum. A custom solution involving a specially designed very rigid frame is required to overcome these difficulties.

Here, we sought to reduce and to isolate environmental vibrations due to our need for stable measurements and noise reduction in our research projects conducted with probes inside and outside a cryostat. Our laboratory utilizes scanning SQUID microscopes to take images and to map the properties of superconducting materials. SQUID microscopes are sensitive to vibrations, and additional cameras and optical parts attached externally to the cryostat are affected [23]. We first explore the design and setup of our active isolation system, then detail simulations of the displacement of modal frequencies at each major resonance of the system. We then describe our experimental methods and display the results of acceleration measurements in the form of one-third octave velocity spectra and transfer functions. Our efforts to reduce vibrations from outside the cryostat have resulted in a drastic attenuation at frequencies between 1 Hz and several hundred Hz from environmental sources of vibration.

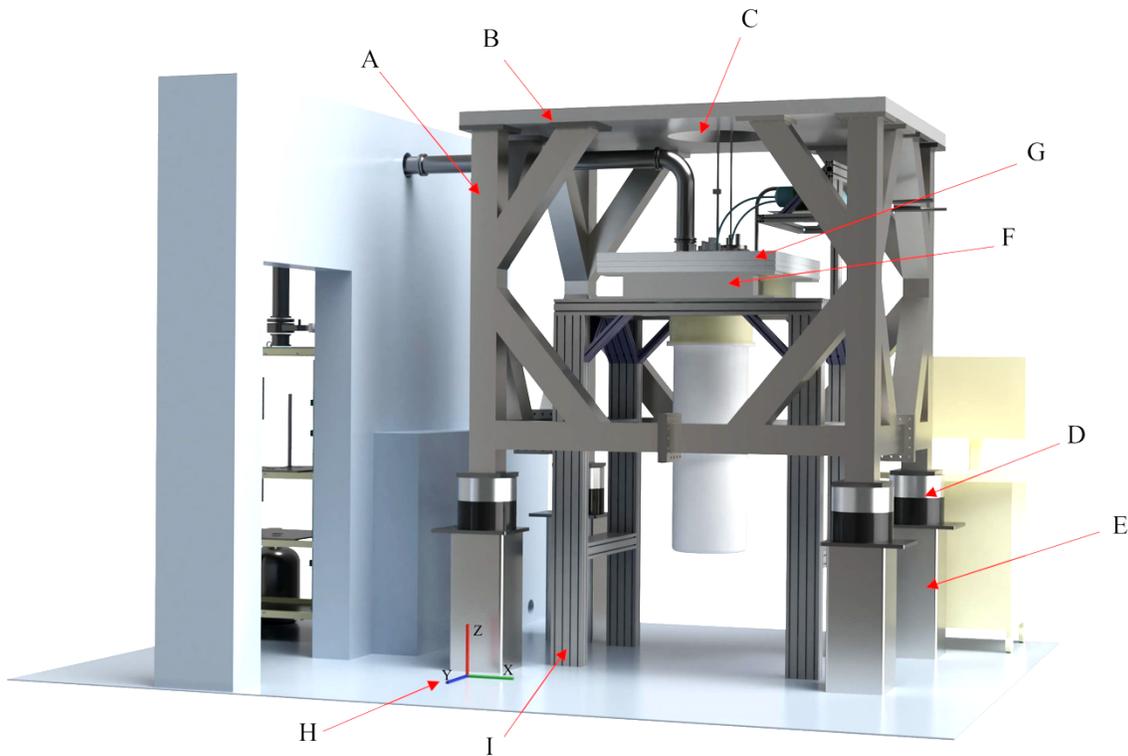

Figure 1 **Computer-aided design (CAD) of vibration isolation for a closed-cycle $^3$He-$^4$He dilution refrigerator.** This CAD image depicts a to-scale rendering of our cryostat setup with the new vibration isolation structure installed. *Isolation Section I:* The tower structure is made of (A) a steel frame with a CleanTop® optical table with a (B) payload plate on top. The optical table has a large opening for installing (C) various experimental probes (optical, mechanical, electrical) that are fed through ports into the cryostat space. (D) TMCTM STACIS® III active isolators upon which the frame sits are placed on (E) very stiff heavy-duty risers made by TMC™ installed on a concrete floor 0.25 m in thickness. The weight of the tower is >2000 kg. *Isolation Section II:* (F) Herzan® AVI series active system suspending (G) 200 kg aluminum plates on which vibrations are measured. The plates are rigidly connected to the cryostat chamber. (H) The axes show the measurement directions (green, X; blue, Y; red, Z). The Herzan® isolators reduce vibrations directed immediately into the cryostat from the floor and (I) the inner aluminum frame. From the top of the payload plate to the floor, the setup measures ~3 m (for photos and proportions, see Figure S 1 ).

# Design of the raised isolation system

Our cryostat must be mounted above the floor at a substantial height to allow access to the instrument both from above and below. This requirement constitutes an engineering challenge that is overcome by our design. Another novel aspect of the present investigation is that our isolation system (Figure 1 and Figure S 1) is suitable for various low-temperature cryostats with closed-loop systems; it can *flexibly* accommodate distinct experimental methodologies [8].

We designed a structure that would surround our experimental setup and actively attenuate vibrations. The outer isolation system consists of an active TMC™ STACIS III ® setup with four isolators (Figure 1, item D) on top of supports (Figure 1, item E), upon which the heavy metal frame (Figure 1, item A) sits. Measurement tools that require stabilization and support, such as optics and cameras, are easily attached to the payload table (Figure 1, item B) and run to the cryostat through the circular opening that is 0.5 m in diameter (Figure 1, item C). Environmental vibrations are attenuated in the active supports before reaching the metal frame and the supported devices. Our inner system has a standard BlueFors LD400 structure (Figure 1, item I) paired with Herzan™ active isolators (Figure 1, item F) supporting the plate that holds the refrigerator (Figure 1, item G). These inner system isolators have an effect similar to that of the STACIS ® isolators in the outer system, but are located at the top of the inner aluminum frame and directly support the weight of the cryostat. This design is adaptable: the passive system can be scaled and designed based on the individual setup. The main difference between the STACIS and the Herzan isolation solutions is that the former produces the effect of a very stiff spring system, while the latter produces the effect of a very soft spring system, by design. In addition, the Herzan system has some ability to counteract driven vibrations in the cryostat itself. The unifying property is that both types of isolators are effective at very low frequencies (isolation starts at ~1-2 Hz; Results and Discussion).

Another way to understand the idea underlying our design of the tower with the optical table (Section I) is that it imitates a large-mass concrete ceiling from which certain parts of the cryostat and the measurement probes could be suspended. For this new artificial ceiling to be efficient, the effect of stiff springs must be achieved with the STACIS isolators in combination with the steel tower. Note that actively driven components, such as the heavy-duty high-pressure compressor hoses, should not be *rigidly* connected to the tower, as this strategy can produce undesired responses from the STACIS isolators themselves. However, relatively weakly coupled (flexible coupling) peripheral connections, such as the $^3$He lines, can indeed be efficiently suspended from the vibration isolation tower.

The design of Isolation Section I is based on both engineering principles and modal frequency finite element analyses to estimate resonances (Figure 2). The structure has resonances within the range of attenuated frequencies by the active system, which should be taken into account when evaluating the system's performance.

Isolation Section II is designed to prevent vibrations from propagating through the inner aluminum frame and also to mitigate vibrations from the pulse-tube and the compressor: components propagating internal vibrations, such as pipes connected to the refrigerator from the compressor, have their vibrations attenuated by the active Herzan linear isolators and passive flexible connections (bellows). However, it is also important that the cross-talk between the two isolation sections is minimized to avoid parasitic feedback, achieved through only weak and flexible links between the two sections and by the use of different types of active isolators.

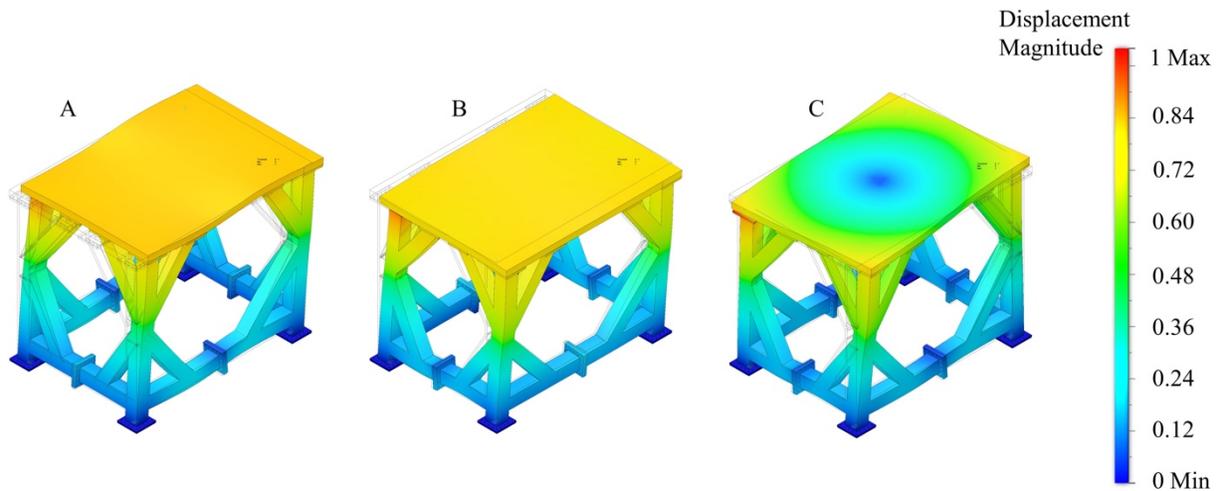

Figure 2 **Simulations of relative displacement of the isolation tower at resonance frequencies for our structural design of a steel tower with a payload plate.** Shown are the three lowest modes; higher modes are small enough in amplitude to ignore (data not shown). The frequencies of the modes are (A) 41.76 Hz, (B) 47.03 Hz, and (C) 70.72 Hz. Vibration magnitude is normalized based on the maximum total (absolute value) displacement, with a scale in the bottom right of each mode. This simulation neglects the fact that the top plate is an optical table with a proprietary (CleanTop®) structure and mechanical properties; instead, we performed calculations with a solid steel slab, which may result in somewhat different frequencies in the simulations compared to the experiment. The images contain exaggerated displacements for visualization. The strong vibrations in the payload plate reflect the most influential vibrational modes within our frequency range. These vibrations are minimized in the real system via an optical table with enhanced mechanical properties rather than a solid metal slab and our active isolation system.

## Results and Discussion

Initial vibration measurements were carried out after installation of the frame using a Wilcoxon 731A seismic accelerometer and P31 amplifier. Measurements were done at various times of day, but all within regular (noisy) business hours. Measurements were taken on the ground for baseline vibration levels, on the payload plate (Figure 1, item B), and on the aluminum plate (Figure 1, item G), with the compressor both on and off. Each set of payload and aluminum plate data is accompanied with its own ground measurements. In all cases, measurements were taken in the X, Y, and Z directions (Figure 1, item H). We expected structural vibrations without an active system to be greater than floor vibrations in general, including frequencies captured in our simulations (Figure 2), while we expected the opposite attenuated behavior with the active system engaged. Details on data analysis and visualization appear in the Supplemental Information.

Figure 3 depicts measurements taken on our payload plate. Transfer functions are presented in dB units to give the magnitude of the output as a function of input frequency, and one-third octave spectra (Figure 3, left) show summed velocity power spectra in the one-third octave bins [24]. This presentation of the data negates artifacts resulting from narrow bandwidth and highlights the power of vibrations in a standardized form [25], whereas the narrowband spectra can be seen in the Supplemental Information. Ground measurements were relatively quiet below 10 Hz, but increased in power substantially to VC-G levels by ~20 dB (Figure 3, left). Payload measurements without active isolation were ~6 dB higher than ground measurements, reaching upward of 20 dB larger before 20 Hz (Figure 3, left). Payload measurements with active isolation experienced vibrations that were substantially lower in power than in the ground and system-off measurements (Figure 3, left). In both the X and Y positions, below 10 Hz this change constituted a 10-dB reduction from ground measurements and a 20-dB reduction from system-off measurements; above 10 Hz, the change was >20 dB of reduction from ground measurements and system-off measurements (Figure 3, left). The Z position showed >40 dB of reduction above 10 Hz (Figure 3), indicating that the isolation system has a larger effect in this direction. Similar conclusions arise from evaluation of the transfer functions (Figure 3, right), verifying the validity of our analysis methods and revealing that the active system and the isolation structure are effective at reducing vibrations on the payload plate from the environment below 200 Hz. Note that the transfer function indicates that our frame amplifies most vibrations, as vibration magnitude is above 0 dB (Figure 3, right). However, the active system fully mitigates this amplification (Figure 3, right), and does not amplify vibrations for the critical given range of frequencies, as higher than a couple hundred Hertz frequencies can be handled by additional simple room-temperature or cryogenic spring-based elements—the higher frequencies do not require *active* isolation solutions.

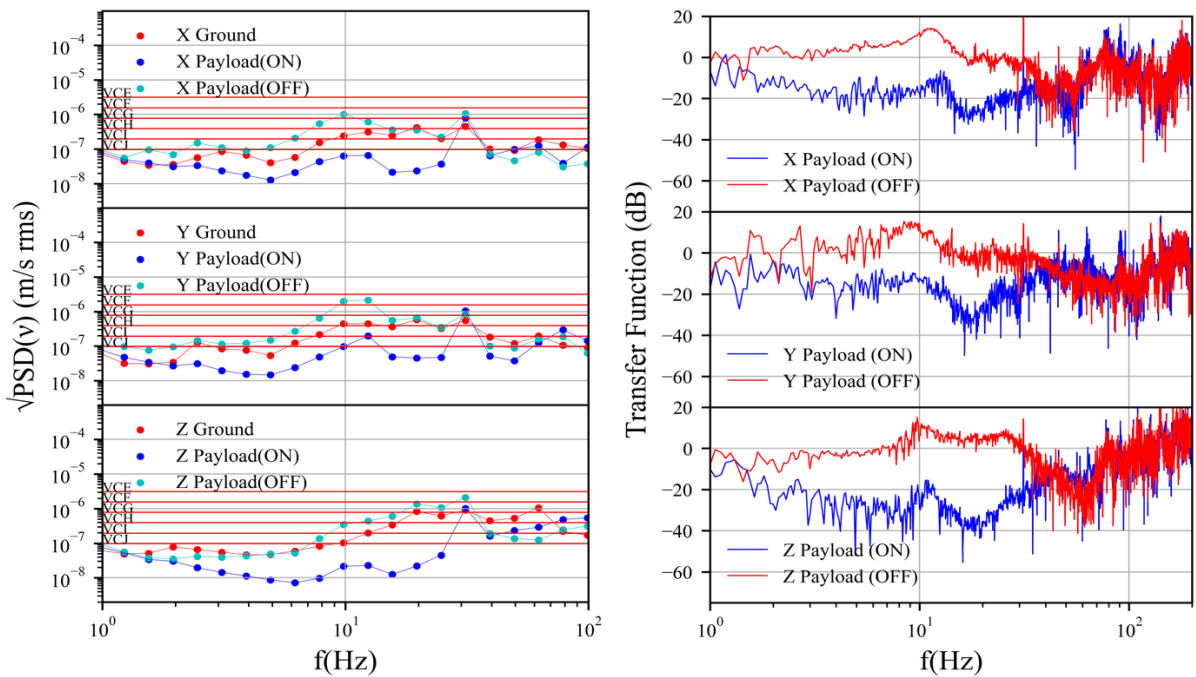

Figure 3 **Strong vibration isolation on the payload plate of at least 20 dB within 2 Hz and 40 Hz from environmental vibrations on the payload plate in all directions.** Left: Octave magnitude of the measured vibrations, in velocity units. Each data point is the sum of the power within each 1/3 octave frequency bin. The red horizontal lines across the figure denote standard levels of vibrations based on the VC criterion [24,25]. Right: Velocity transfer functions calculated using the Amplitude Spectral Density functions of the velocity data. Measurements labeled Ground are taken on the floor, while OFF and ON labels describe the state of the active isolation system. Payload measurements are taken on the CleanTop ® optical table at the top of the tower structure shown in Figure 1. Together, these data reveal the significant effect of the structure's active system on vibrations between ~1 Hz and ~200 Hz, attenuating at structural resonances seen at ~10 Hz and ~20 Hz. For the X and Y directions, this attenuation reaches upward of 30 dB, while Z attenuation reaches values as substantial as 40 dB within the measured frequency range. There is little difference in vibrations after 50 Hz for ON, OFF, and Ground measurements; these (both the floor and the payload) vibrations are already of low magnitude (VC-H or lower), making this area less critical to isolate, in our case. Data were taken with a Low Pass Filter of 450 Hz as well, limiting the value of measurements for frequencies >200 Hz based on the Nyquist Frequency.

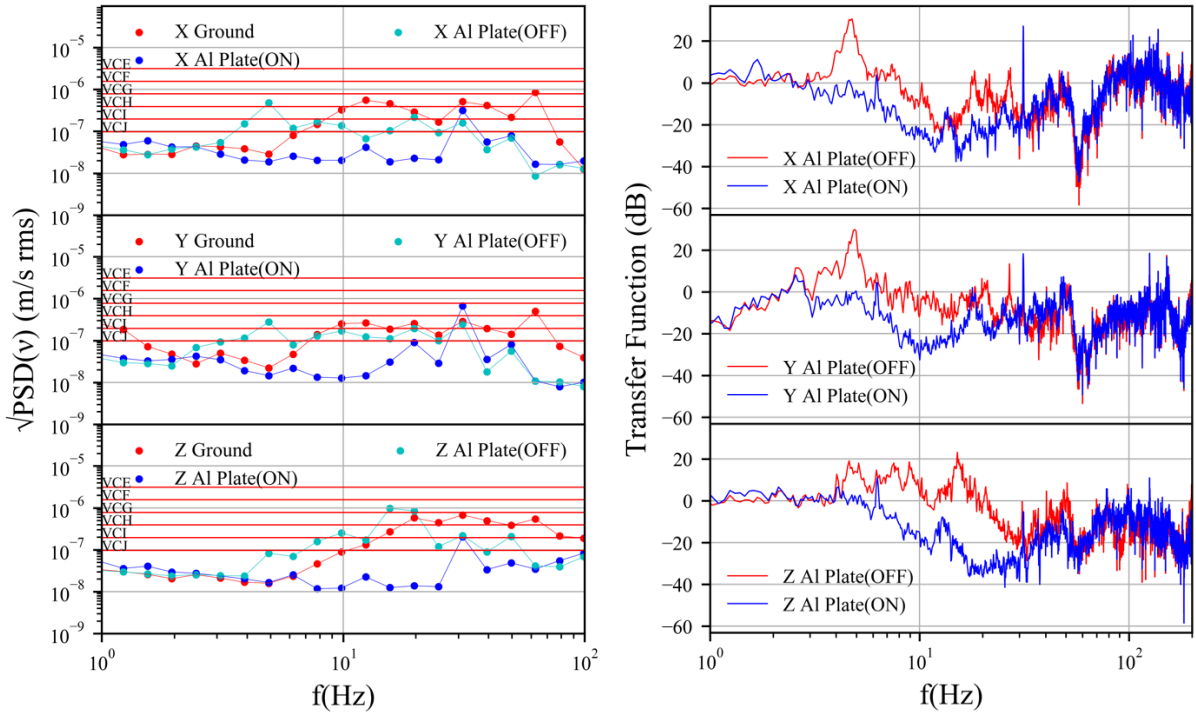

Figure 4 **Vibration measurements taken on the aluminum plate holding the cryostat with the compressor off.** Left: Octave magnitude of the measured vibrations in the aluminum plate, presented in velocity units. Right: Velocity transfer functions. Measurements labeled Ground were taken on the floor, while OFF and ON labels describe the state of the active isolation system. Aluminum Plate measurements were taken on the aluminum plate supporting the cryostat and isolated by the Herzan AVI series active system (Figure 1, item F). Vibrations are attenuated after the nominal ground vibration level ~2 Hz up to 200 Hz. X and Y directions show 20 dB of vibration in this range, with both transfer functions following similar behavior. The Z direction exhibits 30-40 dB of attenuation between 10 Hz and 30 Hz. All three conditions manifest the same peak in vibrations between 30 Hz and 40 Hz, as well as another resonance

around 6 Hz. Our data show at least 20 dB of attenuation for all directions of vibrations in the aluminum plate, highlighting this system's effectiveness in isolating our setup from the environment.

We also measured vibrations on aluminum plates with the compressor off, presenting them in one-third octave (Figure 4, left) and transfer functions (Figure 4, right). Corresponding narrowband spectra are in the Supplemental Information. The ground vibrations were very quiet, <6 Hz, before increasing to VC-I and VC-H levels around 10 Hz (Figure 4). OFF measurement vibrations peaked in the XY directions at around 4 Hz (Figure 4), likely a resonance of the Herzan isolators when inactive. Plate vibrations for the OFF measurements began to increase at lower frequencies, with attenuation of ~20 dB in the ON measurement vibrations (Figure 4, right). For all three directions, the vibration velocity was well below VC-J levels and stayed flat at ~$10^{-8}$ m/s rms up until the peak around 30 Hz and a series of resonances from 30-60 Hz (Figure 4, left). The peaks at 30 Hz and 60 Hz were apparent on both the payload plate (Figure 3, right) and the aluminum plate (Figure 4, right); because these structures are different and have different vibrations profiles between 30 Hz and 60 Hz, these peaks may be the result of electronic noise or feedback. We again detected a more significant effect on the vibrations data in the Z direction, clearly evident in the transfer functions (Figure 4, right) that mirror the octave data (Figure 4, left). Given the behavior of the OFF vibrations, our results indicate that the active system attenuates vibrations from the environment at least up until ~200 Hz on the aluminum plates holding the refrigerator, while reducing the amplification of the frame at most frequencies. Thus, vibrations resulting from the environment are attenuated before they reach the cryostat, negating the influence of a noisy environment on the system—apart from multiple resonance peaks seen in the curves. These resonances appear consistently around 6 Hz, between 10-20 Hz, and between 30-60 Hz.

Narrowband data, including measurements with the compressor on, are reported in the Supplemental Information (Figure S 2-Figure S 5). These data gathered with the compressor ON do show typical levels of vibrations induced by the pulse tube and the compressor. These internal vibrations should be mitigated by cryogenic isolation stages [2,18–21], as our design mainly focuses on external vibrations.

# Summary and Conclusions

Environmental vibrations are a large source of noise in ultra-cold, precise-measurement systems. Proper isolation of the system is key to reducing this noise and to obtaining the precision necessary for technologies requiring such precision. Here, we effectively isolated our cryostat by utilizing a large custom-built support tower and two sections of active isolation system in conjunction with one another. The design of our system was tested through simulations (Figure 2), revealing only three resonance frequencies in the payload (Figure 1) at ~30 Hz, ~60 Hz, and ~90 Hz. With the active system on, the payload experienced a reduction of ~20 dB from ground vibrations and a 40-dB reduction from OFF vibrations, while the aluminum plate experienced a reduction of ~20 dB from both ground vibrations and OFF vibrations in the ON setting between ~2 Hz and ~200 Hz (Figure 3, Figure 4, and Figure S 2-Figure S 5). The aluminum plate measurements were very promising when the compressor was off. With the active system on, vibrations peaked only around the resonances between 30 Hz and 60 Hz; vibrations below VC-J levels were reduced at all other frequencies between ~1 Hz and ~200 Hz.

The system described here effectively negates environmental vibrations even when mounted at a very large height above the floor, which is generally very challenging for off-the-shelf products due to parasitic resonances. We overcame these difficulties with a novel custom rigid frame and a novel dual-stage active isolation system with various types of isolators, thus avoiding parasitic feedback and crosstalk effects.

While the current implementation was carried out for the BlueFors dilution refrigerator, it has a fairly common overall geometry similar to cryostats by other manufacturers. Therefore, the potential applications of this vibration isolation design and its adaptations are wide ranging, from custom-built systems to commercially produced cryostats (not only BlueFors) containing prebuilt isolators and structures to solve vibration at the factory stage. Paired with internal vibration reduction methods [2,18–21], this approach should advance the construction of new, low-temperature research environments, including for quantum information science.

## Supplemental Information

Supplemental Information includes photographs of the installed system, more information about the procedures and methods, and additional vibrations data.

## Acknowledgments

We thank University of Connecticut's Planning Design and Construction (UPDC) team, the engineering and architectural teams involved, the TMC's engineering team, Whiting-Turner team, and Raymond Celmer from the Physics Department for their help with this project.

## References

[1] G. Batey, A. Casey, M.N. Cuthbert, A.J. Matthews, J. Saunders, and A. Shibahara, New J. Phys. **15**, 113034 (2013).

[2] M. Pelliccione, A. Sciambi, J. Bartel, A.J. Keller, and D. Goldhaber-Gordon, Rev. Sci. Instrum. **84**, 033703 (2013).

[3] M. Hashisaka, Y. Yamauchi, K. Chida, S. Nakamura, K. Kobayashi, and T. Ono, Rev. Sci. Instrum. **80**, 096105 (2009).

[4] D. Schmoranzer, S. Kumar, A. Luck, E. Collin, X. Liu, T. Metcalf, G. Jernigan, and A. Fefferman, J. Low Temp. Phys. **196**, 268 (2018).

[5] I. Sochnikov, A. Shaulov, Y. Yeshurun, G. Logvenov, and I. Bozovic, Nat. Nano **5**, 516 (2010).

[6] I. Sochnikov, A.J. Bestwick, J.R. Williams, T.M. Lippman, I.R. Fisher, D. Goldhaber-Gordon, J.R. Kirtley, and K.A. Moler, Nano Lett. **13**, 3086 (2013).

[7] I. Sochnikov, L. Maier, C.A. Watson, J.R. Kirtley, C. Gould, G. Tkachov, E.M. Hankiewicz, C. Brüne, H. Buhmann, L.W. Molenkamp, and K.A. Moler, Phys. Rev. Lett. **114**, 066801 (2015).

[8] C. Herrera, J. Cerbin, K. Dunnett, A.V. Balatsky, and I. Sochnikov, ArXiv:1808.03739 Cond-Mat (2018).

[9] C. Herrera and I. Sochnikov, ArXiv:1907.01733 Cond-Mat (2019).

[10] R. Kalra, A. Laucht, J.P. Dehollain, D. Bar, S. Freer, S. Simmons, J.T. Muhonen, and A. Morello, Rev. Sci. Instrum. **87**, 073905 (2016).

[11] E. Olivieri, J. Billard, M. De Jesus, A. Juillard, and A. Leder, Nucl. Instrum. Methods Phys. Res. **858**, 73 (2017).

[12] Y. Shperber, N. Vardi, E. Persky, S. Wissberg, M.E. Huber, and B. Kalisky, Rev. Sci. Instrum. **90**, 053702 (2019).

[13] L.B.-V. Horn, Z. Cui, J.R. Kirtley, and K.A. Moler, ArXiv:1812.03215 Cond-Mat (2018).

[14] S. Zhang, D. Huang, and S. Wu, Rev. Sci. Instrum. **87**, 063701 (2016).

[15] J.D. Hackley, D.A. Kislitsyn, D.K. Beaman, S. Ulrich, and G.V. Nazin, Rev. Sci. Instrum. **85**, 103704 (2014).

[16] M. Pelliccione, A. Jenkins, P. Ovartchaiyapong, C. Reetz, E. Emmanouilidou, N. Ni, and A.C. Bleszynski Jayich, Nat. Nanotechnol. **11**, 700 (2016).

[17] F.P. Quacquarelli, J. Puebla, T. Scheler, D. Andres, C. Bödefeld, B. Sipos, C. Dal Savio, A. Bauer, C. Pfleiderer, A. Erb, and K. Karrai, Microsc. Today **23**, 12 (2015).

[18] A.M.J. den Haan, G.H.C.J. Wijts, F. Galli, O. Usenko, G.J.C. van Baarle, D.J. van der Zalm, and T.H. Oosterkamp, Rev. Sci. Instrum. **85**, 035112 (2014).

[19] M. de Wit, G. Welker, K. Heeck, F.M. Buters, H.J. Eerkens, G. Koning, H. van der Meer, D. Bouwmeester, and T.H. Oosterkamp, Rev. Sci. Instrum. **90**, 015112 (2019).

[20] R. Maisonobe, J. Billard, M.D. Jesus, A. Juillard, D. Misiak, E. Olivieri, S. Sayah, and L. Vagneron, J. Instrum. **13**, T08009 (2018).

[21] C. Lee, H.S. Jo, C.S. Kang, G.B. Kim, I. Kim, S.R. Kim, Y.H. Kim, H.J. Lee, J.H. So, and Y.S. Yoon, J. Instrum. **12**, C02057 (2017).

[22] L. Zhang, T. Miyamachi, T. Tomanić, R. Dehm, and W. Wulfhekel, Rev. Sci. Instrum. **82**, 103702 (2011).

[23] D. Schiessl, J.R. Kirtley, L. Paulius, A.J. Rosenberg, J.C. Palmstrom, R.R. Ullah, C.M. Holland, Y.-K.-K. Fung, M.B. Ketchen, G.W. Gibson, and K.A. Moler, Appl. Phys. Lett. **109**, 232601 (2016).

[24] A. Amick, M. Gendreau, T. Busch, and C. Gordon, in Proc. SPIE 5933 (2005).

[25] C.G. Gordon, in Proc. SPIE 3786 (1999).

# Supplemental Information

## The Experimental System

Our cryostat appears in the center of the photographs in Figure S 1. The outside larger frame (gray) standing on the black steel risers is the steel tower of Isolation Section I with a payload plate (CleanTop®). The inner (silver) aluminum structure holding our BlueFors LD-400 is standard with the refrigerator, except for the Herzan isolators at the top between the structure and the aluminum plate.

This system is "Level I" package in terms of the vibration isolation provided by Bluefors with the system, which includes aluminum plates, one flexible bellow for the pulse-tube, two flexible bellows at the gas-handling system, and copper braids at the pulse tube. In the Level I package, the pulse-tube can be suspended from the actual building ceiling. The ceiling is reached through the large round opening in the middle of our optical table. The pulse-tube mount is not shown in the current photos, but it is a fairly standard tactic for isolating the pulse-tube in many cryostats and can be easily reproduced by other research groups.

The relevant pieces of our setup are labeled in Figure 1 of the main text; all of these pieces also appear in Figure S 1. With this vibration isolation system, our cryostat achieves the ultra-low temperatures necessary for scanning SQUID microscopy [8].

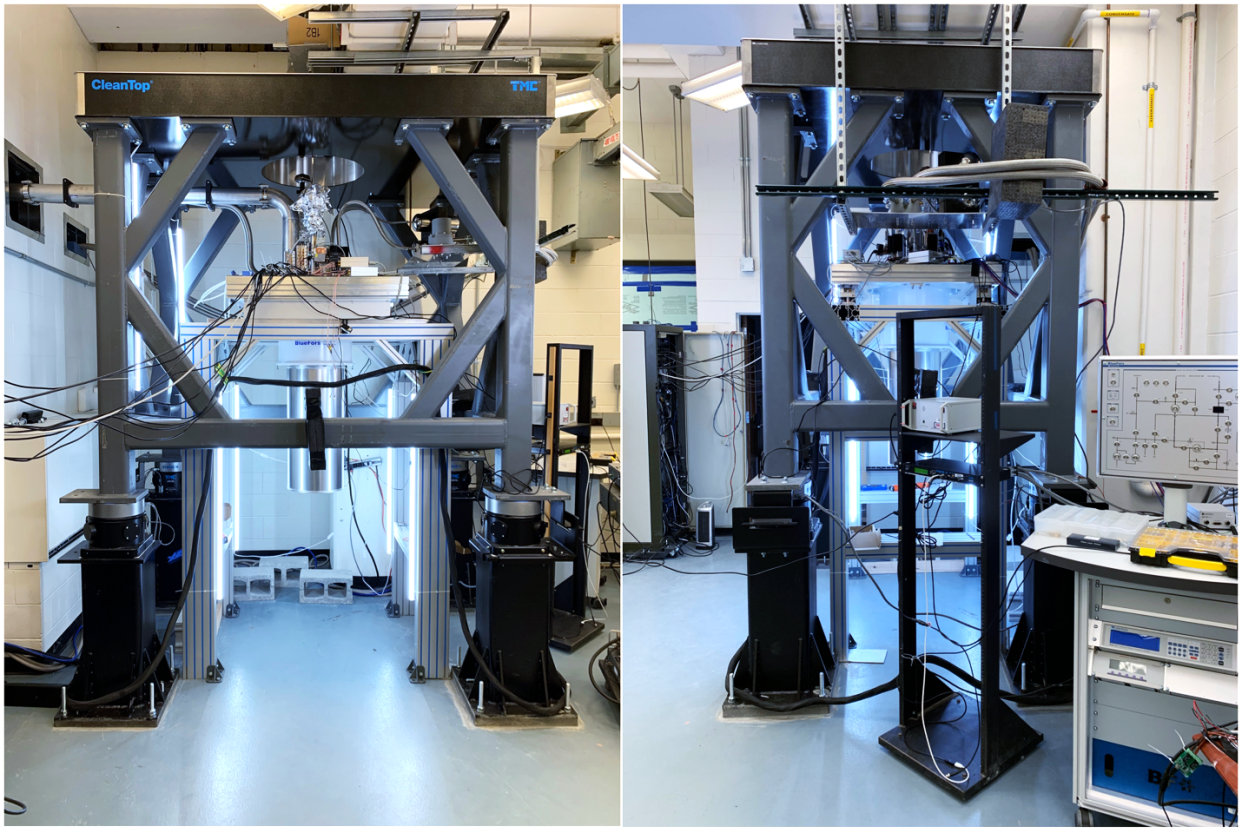

Figure S 1 **Photographs of our cryostat setup and vibration isolation system.** These photographs depict the proportions of the isolation system in a realistic setting. Many details of the setup, including the experimental wiring, vacuum tubing, and electronics, are not detailed in a rendered design (Figure 1). Isolation Section I (the outer section) of our system consists of the optical table on top of the steel frame, which is installed on the TMC™ STACIS III ® active isolators. This frame serves as a stabilization platform for some of our optical setup and staging for other measurement probes. Isolation Section II (the inner section) consists of the aluminum frame, the Herzan AVI isolators, the aluminum plate holding the refrigerator and electronic connectors, and the refrigerator itself. This inner section reduces external vibrations using the Herzan active isolators.

## Calculations

Our power spectral density (PSD) function programmed in MatLab takes an input signal of time-trace voltage from a data acquisition (DAQ) card and produces an output PSD spectrum in units of $V^2$/Hz. Our Wilcoxon P31 amplifier has a gain of 10 V/V and a calibrated sensitivity of 11 V/$g$, resulting in a net gain of 110 V/$g$ on output data. Dividing our PSD spectrum by (110 V/$g$)$^2$ and taking the square root of all values yields acceleration units of $g$/√Hz. Converting these acceleration units to the velocity units needed for the 1/3 octave graphs in Figure 3 and Figure 4 is accomplished from the Fast Fourier Transform of the time data by using the formula $a = v\omega$. Multiplying by 9.81 for $g$ gives us the acceleration amplitude spectral density in m/s$^2$/√Hz. Dividing the acceleration spectra by the angular frequency ω, or 2πf, where f is the measurements frequency, gives the desired velocity amplitude spectral density in units of m/s/√Hz. These two units are seen in the Web Links section of this Supplemental Information for the data. The link to Python™ code used in this investigation also appears there.

1/3 octave calculations were carried out by summing the power spectra within regular 1/3 octave bands. We used 1000 Hz as our reference center frequency, $f_0$. The upper and lower limits of the bands were $f_0 \times 2^{1/6}$ and $f_0/2^{1/6}$, respectively. We summed the power within each band to obtain magnitudes for the 1/3 octave bands. For a reference on 1/3 octave calculations, see the Web Links section of this Supplemental Information. For a description of the VC criterion and its significance, see Ref [24,25].

Our transfer functions were calculated by taking the square root of the output PSD divided by the input PSD (the structural measurement spectra and ground spectra, respectively). The conversion to decibels from power is:

$20 \times log_{10}(\sqrt{\frac{PSD_{out}}{PSD_{in}}})$.

We used a definition in which the square root of the PSD is the Amplitude Spectral Density function. This definition is used for all our transfer functions, 1/3 octave narrowband velocities, and acceleration vibration measurements in Figure 3, Figure 4, Figure S 2, Figure S 3, Figure S 4, and Figure S 5.

Extended Data

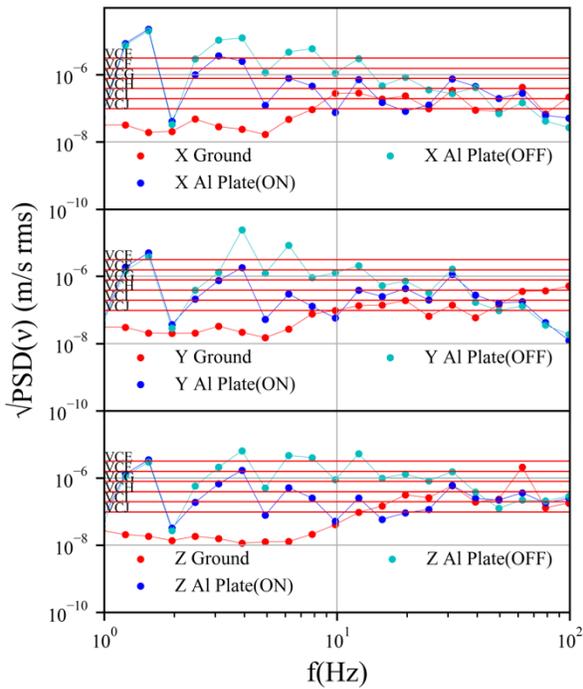

Figure S 2 **Vibration velocities measured on the aluminum plate with the coldhead and the compressor ON reveal active vibration attenuation despite the prominence of compressor-driving frequencies.** The 1/3 octave bins, in velocity rms units, reveal the summed power of the vibrations. This visualization is useful in showing that VC-G levels of vibrations are achievable between ~2 Hz and 200 Hz despite the increase in magnitude from compressor vibrations. We cannot display a transfer function for these measuremets due to the nature of the vibrations: the compressor vibrations are internally driven, so a transfer function relative to the environmental vibrations is meaningless. Reasonable attenuation levels are still achievable with the compressor ON. While driving frequencies remain an issue with our current system, several works have addressed these vibration with cryogenic internal isolation solutions [2,18–21]. Our data show vibration spikes attributable to compressor driving frequencies and effective isolation on the aluminum plate with the compressor ON, which yields effective isolation within the cryostat even in the presence of driven vibrations.

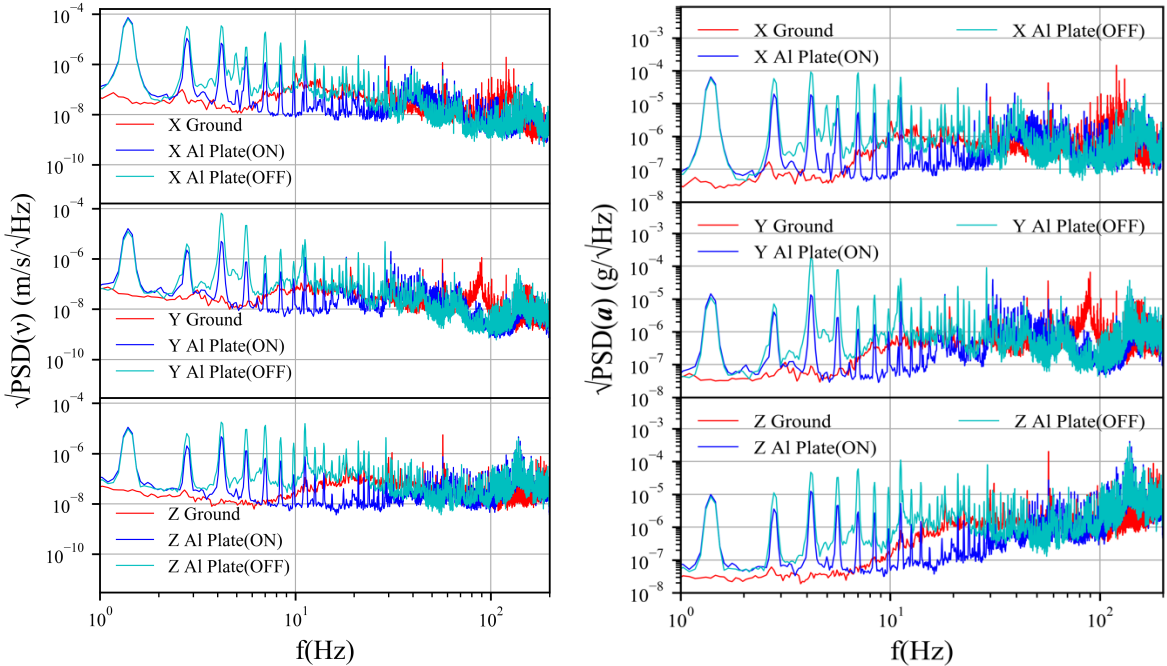

Figure S 3 **Narrowband velocity (left) and acceleration data (right) for the aluminum plate with the compressor on, showing data used in Figure S 1 and the prominence of compressor-driven frequencies.** Due to the compressor and the pulse tube being turned on, these data show obvious peaks at the harmonics of the driving frequency (1.4 Hz) of the pulse tube. These vibrations are present in all directions. Left: Velocity narrowband data demonstrate that the active system attenuates these driven vibrations in all positions. Although the magnitudes of the peaks contribute to a larger overall power in the vibrations, peaks only occur at specific frequencies. Right: Matching acceleration data for aluminum plate measurements with the compressor ON. The frequency spikes are present at the same frequencies as in the narrowband velocity data, and similar attenuation effects appear. The active system attenuates even these driven vibrations superimposed with the external (environmental) vibrations, which combined with internal solutions [2,18–21] leads to a satisfactory level of internal vibrations on cold plates [8].

Figure S 2 and Figure S 3 contain measurements on the aluminum plate with the compressor ON. They show prominent peaks due to internal compressor vibrations, where specific frequencies exhibit large vibration magnitudes. Bins between each of these peaks in the narrowband display attenuation at levels similar to those in Figure 4. Another use comparison would be to the spectra of Figure S 5. The spectra in Figure S 3 clearly reflect those in Figure S 5, but with the superimposed compressor-driven vibrations. Despite the strength of these driving frequencies, the active system still reduces vibrations in the frame, with levels at nearly all frequencies lower than the OFF measurements (**Figure S 3**). The ON measurements still reveal a noticeably lower power in vibrations even in most compressor mode peaks by at least an order of magnitude (**Figure S 2**). We left out a transfer function for these data because we cannot reasonably make one between the structure and the ground. The internal vibrations are not present in the ground, and therefore the input signal for the transfer function would have to include both ground and compressor vibrations.

There are some caveats to the analysis presented here. The intent of our installed system is to reduce environmental vibrations, which it does very effectively (see main text). However, all other experiments performed in the lab are conducted while the compressor is ON, causing internal vibrations. Our structure constitutes a first step in reducing these vibrations by stabilizing the experimental probes on the payload plate frame. It is not out of the question that additional passive components could be installed, possibly spring systems, to attenuate vibrations at the specific frequency modes of the compressor.

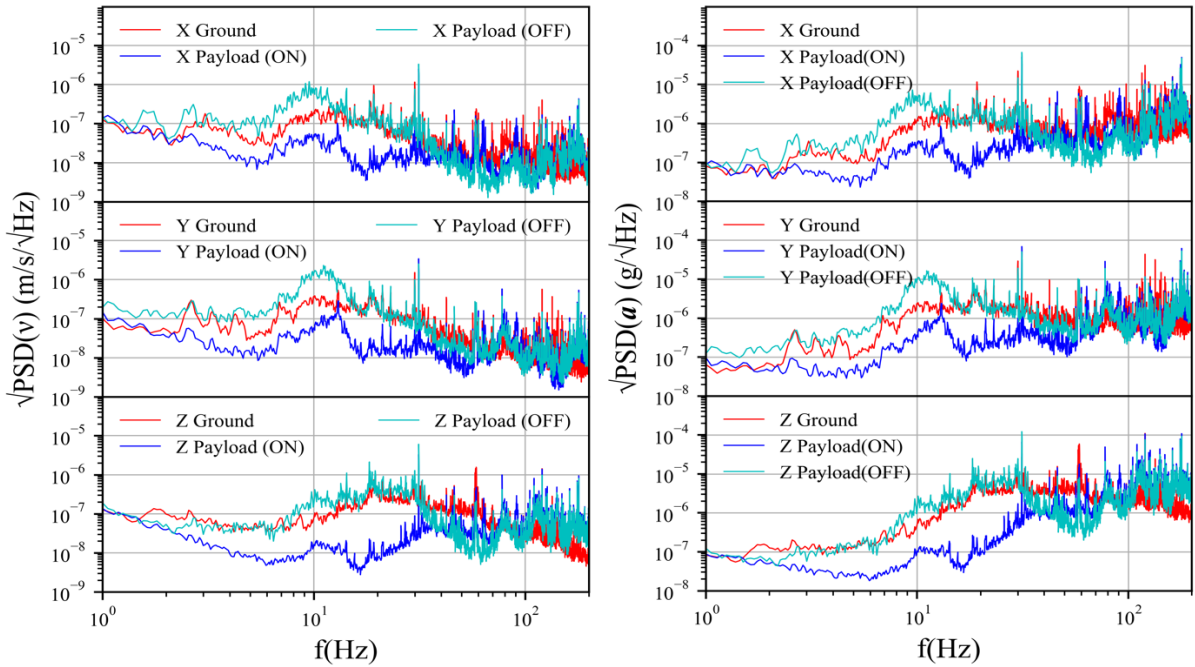

Figure S 4 **Velocity (left) and acceleration (right) PSD spectra for payload plate measurements, highlighting the effectiveness of the active system.** Left: The velocity narrowband data arise from the conversion from acceleration to velocity detailed in the Calculations section of this Supplemental Information. These are the exact spectra from which the 1/3 octave magnitudes are found within their respective bins. The narrowband data reveal more detailed information about spikes in vibration magnitude, but follow the same overall trend as the 1/3 octave plots of the main text. The active system measurements indicate ~10 dB and ~20 dB of attenuation from the floor and OFF measurements, respectively. Isolation is again most effective in the Z direction. Right: Acceleration data are directly conversed from spectral voltage measurements, as described in the Calculations section. Because acceleration and velocity are related, their spectra are very similar. There is a clear peak between 10 Hz and 20 Hz, likely due to some sort of resonance. Levels of attenuation are similar to those in the velocity figure (left panel). These graphs present the narrowband data for the payload measurements reported in the main text and have similar levels of attenuation with more detailed information about specific frequency strengths for each individual measurement, as opposed to the binned data (octaves) and relationship-dependent presentation (transfer functions).

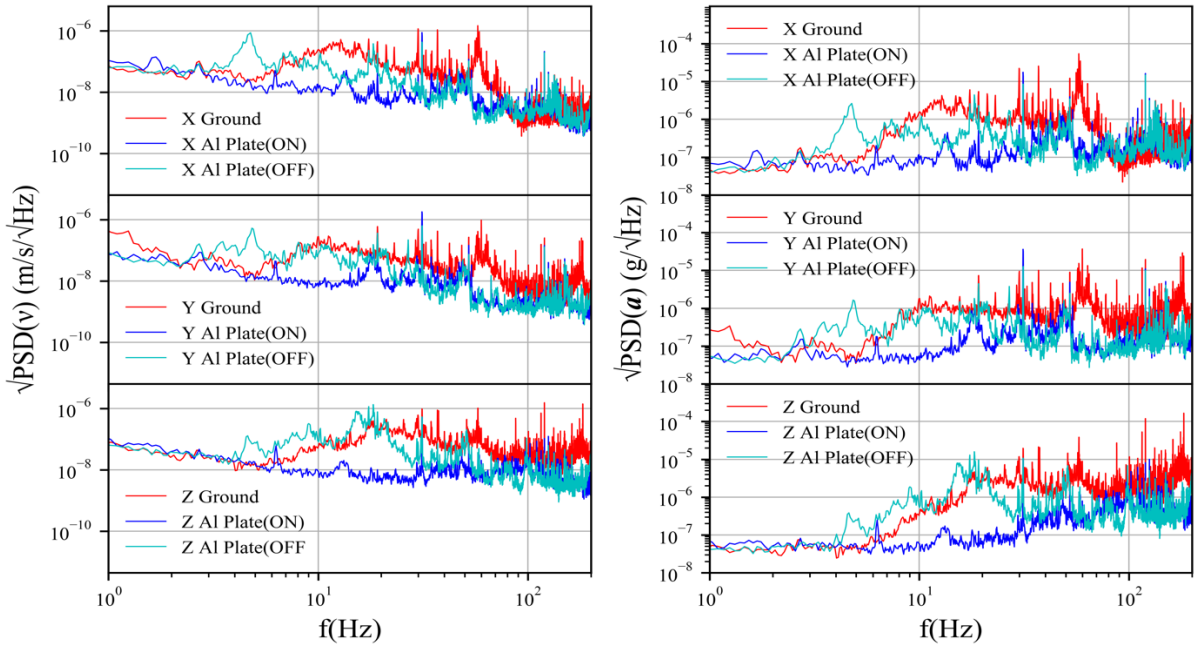

Figure S 5 **Narrowband velocity (left) and acceleration (right) data for the aluminum plate with the compressor OFF, including data used in Figure 4 and the effectiveness of the active system.** Left: Velocity narrowband data demonstrate that the active system attenuates vibrations between ~2 Hz and 200 Hz in all positions. There is a clear spike in vibration magnitude around 5 Hz without the active system ON, which is likely a resonance in the system's architecture when it is OFF. Attenuation is strongest in the Z direction, but all directions reach 20 dB of reduction with the active system. Right: Matching acceleration data for aluminum plate measurements with the compressor OFF. There are strong similarities between the velocity and acceleration data due to their relationship, but acceleration is derived directly from the raw data. Conversions to velocity and acceleration are described in the Calculations section of this Supplemental Information.

Figure S 4 and Figure S 5 contain the velocity and acceleration PSD spectra used to create the figures in the main text. The acceleration behavior is reflected in the velocity spectra because it is a direct conversion, 1/3 octave figures were created using the same velocity spectra, and the transfer functions were made using the raw output PSD but have similar behavior to both velocity and acceleration spectra. The most valuable aspect of these plots is being able to examine the narrowband behavior of the spectra before summing in bins, in order to determine which frequencies contribute the most or exhibit abnormal power. For example, the payload manifests what looks like a resonance frequency at ~10 Hz, which may be an artifact of the payload plate architecture or active system architecture modes (when off). Conclusions made previously in this Supplemental Information about the spectra with the compressor ON and OFF can also be made here, as the underlying spectra (the base-line) in Figure S 3 without the superimposed modes of the compressor are identical to the spectra in Figure S 5.

The code used for conversions between PSD units, transfer functions, and octaves appears in the Web Links section of this Supplemental Information alongside the data used in our calculations.

# Improvements

The effectiveness of our active system is realized via a two-sections design: the payload plate with the strongest vibrational resonances, and the aluminum plate that holds the refrigerator. Our simulations in Figure 2 predicted correctly that the resonances in our steel structure reside at frequencies of 30-90 Hz. Our payload plate is associated with ~30 Hz, ~60 Hz, and 90 Hz resonance peaks in all directions, while other peaks show up in separate directions (Figure 3). The Y and Z directions had a peak at ~80 Hz that was nonexistent in the X direction, while the X and Z directions had a peak at ~40 Hz that was absent from the Y direction (Figure 3). Figure S 4 contains narrowband data for the payload plate for further clarity. In all cases, vibrations that are not completely attenuated by our active system (that show magnitudes ~0 dB in the transfer function) are due to resonances. The 30 Hz and 60 Hz peaks may simply be artifacts due to noise picked up from the electrical grid's 60 Hz signal as the first harmonic and the subharmonic. Nonetheless, the resonance vibrations were overall lower in our plate measurements (Figure 3), suggesting that our system is effective at reducing vibrations throughout the frame but may be mildly susceptible to these resonance modes.

For the aluminum plate there are spikes that are likely due to structural resonances in the aluminum frame (Figure 4). This observation does not detract from the attenuation in the aluminum plates. In the Y and Z directions nearly all frequencies show vibration levels <0 dB for the ON data, except for a couple of peaks for each condition (Figure 4). The X direction exhibits similar levels of attenuation, aside from a small section around 100 Hz that is likely just noise due to division by very small values of the ground vibrations when the transfer function is calculated (Figure 4, right). The main point here is that despite the existence of resonances in the frame, all but the 6 Hz and 30 Hz peaks are attenuated to levels below ground vibration. The narrowband behavior of these vibrations can be seen in Figure S 5, which shows the same small peaks in this range as in Figure 4.

We are planning future work to suppress these resonances. There are several methods that we can apply to the structure to help suppress resonances. For example, the frames are hollow; a filling of a heavy material or one that can help absorb vibrations (such as lead shot, sand, or polymers) could mitigate resonance [26]. Securing connections between parts of the frame with welding instead of bolts is another planned improvement, which we are planning to implement when the cryostat is moved to its new permanent location. It is also possible that our structure is picking up acoustic vibrations, such as from air conditioning, suggesting the addition of fairly straightforward acoustic isolation methods using absorbing materials [27]. Even accounting for resonances, the transfer functions in Figure 3 and Figure 4 indicate that our active system is very effective at reducing vibrations across relevant spectra to, at the very least, the baseline magnitude of a vibration-quiet building.

Last, we detail how to connect measurement probes to Sections I and II. The optics of our system essentially consists of off-the-shelf 30-mm or 60-mm standard cage systems and their lens tubes from Thorlabs Inc. (New Jersey, USA). The optics is connected, at one end, to a standard vacuum port with a fused silica window; at the other end, it is connected directly to the optical table of Section I. The connection through the cage system is sufficiently flexible that it causes no apparent performance issues in terms of decoupling Section I and Section II of the isolation system. For a distinct, previous application, with a mechanically driven strain cell [8], the lead axis was fed through a Rigaku (New Hampshire, USA) ferrofluid rotary motion feedthrough (Part Number FD-KF-0500-LC). It was then coupled to a low-speed high-torque motor directly mounted on the optical top of the tower (Section I) through a standard off-the-shelf flexible shaft connection to limit the propagation of vibrations from Section I to Section II. The $^3$He return line was simply suspended from the optical table using a standard flexible strap. For other custom needs, the connections between Sections I and II should be essentially as flexible as the application allows, but there is no need for extremely weak coupling.

## Web Links

The raw data and Python workbook used to analyze the raw data are stored in this GitHub repository:

https://github.com/JSwooty/Vibration-Calculation-Sample-Code.git

## Supplemental References

[26] S. Koch, F. Duvigneau, R. Orszulik, U. Gabbert, and E. Woschke, J. Sound Vib. **393**, 30 (2017).
[27] R. Fleury, D.L. Sounas, C.F. Sieck, M.R. Haberman, and A. Alù, Science **343**, 516 (2014)